\documentclass[12pt]{iopart}

\usepackage{iopams}  

\usepackage{cite}
\usepackage{graphicx}
\usepackage{indentfirst}
\usepackage{braket}
\usepackage{float}

\usepackage{epstopdf}
\usepackage{CJK}
\usepackage{esint}
\usepackage{color}
\usepackage[T1]{fontenc}
\usepackage{subfigure}
\usepackage{letltxmacro}
\LetLtxMacro{\ORIGselectlanguage}{\selectlanguage}
\makeatletter
\DeclareRobustCommand{\selectlanguage}[1]{%
  \@ifundefined{alias@\string#1}
    {\ORIGselectlanguage{#1}}
    {\begingroup\edef\x{\endgroup
       \noexpand\ORIGselectlanguage{\@nameuse{alias@#1}}}\x}%
}
\newcommand{\definelanguagealias}[2]{%
  \@namedef{alias@#1}{#2}%
}
\makeatother

\definelanguagealias{en}{english}
\usepackage{amsfonts}
\usepackage{footmisc}
\usepackage{scrextend}
\usepackage{multirow}
\usepackage{cleveref}
\usepackage[english]{babel}
\usepackage{url}

\newcommand{\qed}{\nobreak \ifvmode \relax \else
      \ifdim\lastskip<1.5em \hskip-\lastskip
      \hskip1.5em plus0em minus0.5em \fi \nobreak
      \vrule height0.75em width0.5em depth0.25em\fi}
      
\def\be{\begin{equation}}
\def\ee{\end{equation}}
\def\ba{\begin{eqnarray}}
\def\ea{\end{eqnarray}}

\begin{document}

\title[]{Entanglement-enhanced Neyman-Pearson target detection using quantum illumination}

\author{Quntao Zhuang$^{1,2}$, Zheshen Zhang$^1$ and Jeffrey H Shapiro$^1$}
\address{$^1$Research Laboratory of Electronics, Massachusetts Institute of Technology, Cambridge, Massachusetts 02139, USA\\
$^2$Department of Physics, Massachusetts Institute of Technology, Cambridge, Massachusetts 02139, USA}

\ead{quntao@mit.edu}
\vspace{10pt}
\begin{indented}
\item March 7, 2017
\end{indented}

\begin{abstract}
Quantum illumination (QI) provides entanglement-based target detection---in an entanglement-breaking environment---whose performance is significantly better than that of optimum classical-illumination target detection.  QI's performance advantage was established in a Bayesian setting with the target presumed equally likely to be absent or present and error probability employed as the performance metric.   Radar theory, however, eschews that Bayesian approach, preferring the Neyman-Pearson performance criterion to avoid the difficulties of accurately assigning prior probabilities to target absence and presence and appropriate costs to false-alarm and miss errors.   We have recently reported an architecture---based on sum-frequency generation (SFG) and feedforward (FF) processing---for minimum error-probability QI target detection with arbitrary prior probabilities for target absence and presence.  In this paper, we use our results for FF-SFG reception to determine the receiver operating characteristic---detection probability versus false-alarm probability---for optimum QI target detection under the Neyman-Pearson criterion.
\end{abstract}

\section{Introduction}
Entanglement is arguably the premier quantum-mechanical resource for obtaining sensing performance that exceeds limits set by classical physics.  Entanglement, however, is vulnerable to loss and noise arising from environmental interactions.  As a result, the performance advantages of many entanglement-enabled sensing schemes---such as those that rely on frequency-entangled states (see, e.g.,~\cite{Giovannetti2001}), or N00N states (see, e.g.,~\cite{Dowling2008})---vanish as loss and noise increase.  Quantum illumination (QI)~\cite{Sacchi_2005_1,Sacchi_2005_2,Lloyd2008,Tan2008,Lopaeva_2013,Guha2009,Zheshen_15,Barzanjeh_2015}, in contrast, is highly robust against environmental loss and noise. QI utilizes entanglement to beat the performance of the optimum classical-illumination (CI) scheme for detecting the presence of a weakly reflecting-target that is embedded in a very noisy environment, despite QI's initial entanglement being destroyed \emph{before} the target-detection quantum measurement is made. In particular, for equally-likely target absence or presence, Tan \emph{et al}~\cite{Tan2008} showed that QI's error-probability exponent is 6\,dB higher than that of the optimum CI scheme of the same transmitted power.   Tan \emph{et al} obtained their result from the quantum Chernoff bound (QCB)~\cite{Audenaert2007}, which gives no inkling as to what receiver hardware could be used to realize that performance advantage.  Indeed, finding a structured optimum receiver for QI has been a longstanding problem. Guha and Erkmen~\cite{Guha2009} introduced and analyzed the optical parametric amplifier (OPA) receiver, showing that its error-probability exponent for equally-likely target absence or presence is 3\,dB greater than that of optimum CI.  A subsequent experiment~\cite{Zheshen_15}, which implemented the OPA receiver, verified that QI could outperform CI in an entanglement-breaking scenario.  In recent theoretical work~\cite{Zhuang2017}, we showed that sum-frequency generation (SFG) combined with a feedforward (FF) mechanism can achieve QI's full 6\,dB advantage in error-probability exponent for equally-likely target absence or presence.   

Tan \emph{et al}'s~\cite{Tan2008} assumption of equally-likely target absence or presence and use of error probability as a performance metric makes their analysis Bayesian, but Bayesian analysis is \emph{not} the preferred approach for target detection, owing to the difficulty of accurately assigning prior probabilities to target absence and presence and appropriate costs to false-alarm (Type-I) and miss (Type-II) errors.  Instead, radar theory opts for the Neyman-Pearson performance criterion, in which optimum target detection maximizes the detection probability, $P_D \equiv \Pr(\mbox{decide present}\mid\mbox{present})$, subject to a constraint on the false-alarm probability, $P_F \equiv \Pr(\mbox{decide present}\mid \mbox{absent})$.   (The detection probability satisfies $P_D = 1-P_M$, where $P_M \equiv \Pr(\mbox{decide  absent}\mid \mbox{present})$ is the miss probability.)  Spedalieri and Braunstein~\cite{Spedalieri2014} derived the optimum trade-off between the false-alarm and miss-probability error exponents in the asymptotic ($M\rightarrow \infty$) limit of $M$-copy quantum-state discrimination. More recently, Wilde \emph{et al}~\cite{Wilde2016} showed that for fixed false-alarm probability, QI's miss-probability exponent greatly exceeds that of the optimum CI scheme.  For weakly-reflecting targets embedded in high-brightness noise, however, Wilde \emph{et al}'s result only holds when $P_M$ is extremely low, e.g., $P_M \sim10^{-30}$ or lower.  In this paper we use results from our FF-SFG analysis~\cite{Zhuang2017} to obtain the receiver operating characteristic (ROC)---i.e., the trade-off between $P_D$ and $P_F$---for optimum QI target detection, and compare it to the ROCs of QI target detection with OPA reception and optimum CI target detection.

The rest of the paper is organized as follows.  In Sec.~\ref{QIdetection} we describe QI target detection, as introduced by Tan \emph{et al}~\cite{Tan2008}, and contrast two general approaches to multiple-copy, quantum-state discrimination that will help later in understanding why QI target detection with OPA reception is inferior to QI target detection using FF-SFG reception.  Sections~\ref{OPArcvr} and \ref{SFGrcvr} are devoted, respectively, to descriptions of the OPA and FF-SFG receivers for QI target detection, including how they use the returned-signal and stored-idler mode pairs that QI provides to make their decisions as to target absence or presence.  Section~\ref{ROCsec} concludes the paper with a comparison between the ROCs of QI target detection with FF-SFG reception, QI target detection with OPA reception, CI target detection with homodyne detection, and a coherent-state discrimination problem whose performance is the ultimate limit for QI target detection in the $N_S \ll 1$ regime.

\section{Target detection using quantum illumination}
\label{QIdetection}
\begin{figure}[hbt]
\centering
\subfigure[]{
\includegraphics[width=0.45\textwidth]{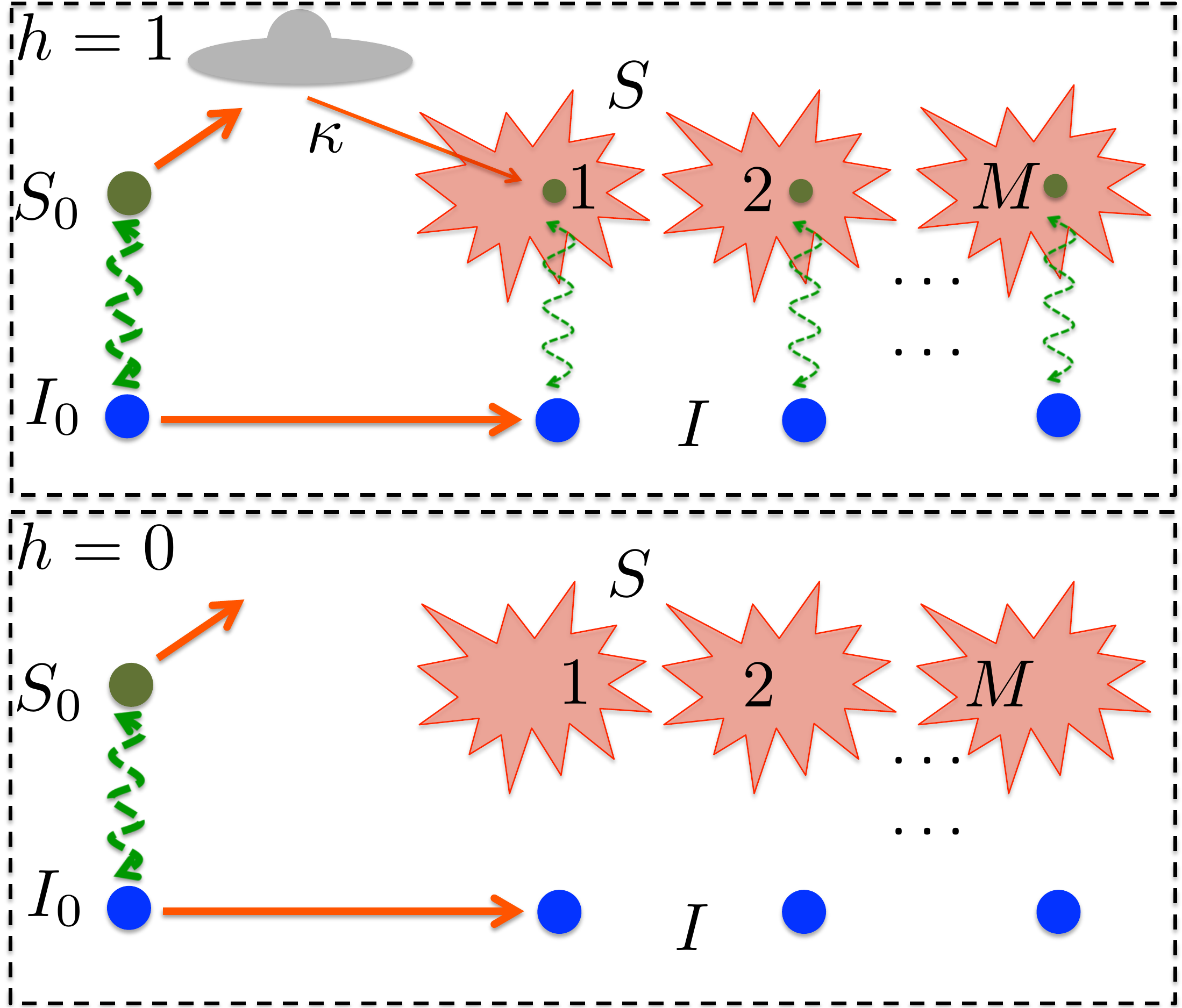}
\label{FIG_schematic}
}
\subfigure[]{
\includegraphics[width=0.45\textwidth]{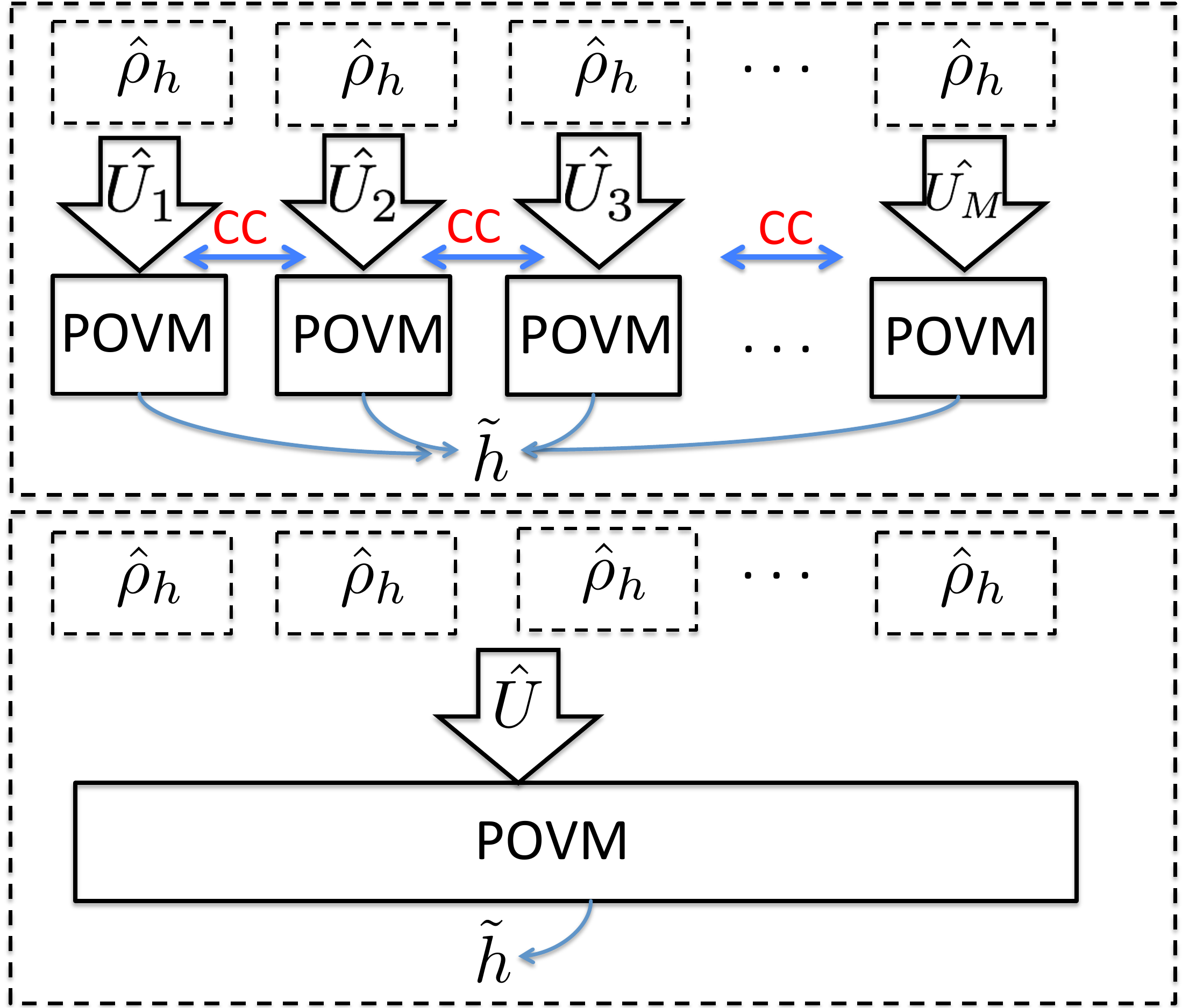}
\label{FIG_LOCC}
}
\caption{(a) Schematic of QI target detection. Upper panel: target present ($h=1$). Lower panel: target absent ($h=0$). The green dashed lines shows the correlation between the signals (green balls) and the idlers (blue balls), with thickness indicating correlation strength.
(b) Measurement schemes. Upper panel: local operations plus classical communication (LOCC) using individual unitary transformations followed by positive operator-valued measurements (POVMs) for each mode pair that may be connected by classical (feedforward or feedback) communication and whose outputs are pooled to reach a final target absence or presence decision ($\tilde{h} = 0$ or 1). Lower panel: collective operation using a unitary transformation $\hat{U}$ operating on all the mode pairs followed by a single POVM to reach a target absence or presence decision.
}
\end{figure}

Figure~\ref{FIG_schematic} is a schematic representation of QI target detection~\cite{Tan2008}. An entanglement source generates $M \gg 1$ independent, identically distributed (iid) signal-idler mode pairs, with photon annihilation operators $\{\hat{c}_{S_{0_m}},\hat{c}_{I_{0_m}} : 1\le m\le M\}$. Each mode pair is in a two-mode squeezed-vacuum state of mean photon number $2N_S \ll 1$. For simplicity, Fig.~\ref{FIG_schematic} shows only a single signal-idler pair $\left(S_0,I_0\right)$. The bold green dashed line denotes the maximum entanglement between $\left(S_0,I_0\right)$. The signal modes (solid green circles) interrogate the region in which the weakly-reflecting target would be located were it present.  The idler modes (solid blue circles) are retained for subsequent joint measurement with noisy signal modes that are returned from the interrogated region. Assuming ideal idler storage, the annihilation operators for the idler modes at the joint measurement are $\{\hat{c}_{I_m} = \hat{c}_{I_{0_m}}: 1\le m \le M\}$. 

Figure~\ref{FIG_schematic}'s upper panel shows that in the presence of a target, i.e., $h=1$, the returned signal modes contain a weak reflection from the target (the small solid green circle) embedded in a bright noise background (the red cloud). The residual signal photons from the transmitter have a weak phase-sensitive cross correlation with the stored idler modes, indicated by the green dashed lines. Thus, when $h=1$ the returned signal modes are described by the annihilation operators $\{\hat{c}_{S_m} = \sqrt{\kappa}\,\hat{c}_{S_{0_m}}+\sqrt{1-\kappa}\,\hat{c}_{N_m}: 1\le m \le M\}$, where $\kappa\ll1$ is the transmitter-to-target-to-receiver transmissivity  and the $\{\hat{c}_{N_m}\}$ are annihilation operators for noise modes, each of which is in a thermal state containing $N_B /(1-\kappa) \gg 1$ photons on average. 

Figure~\ref{FIG_schematic}'s lower panel shows that in the absence of a target, i.e., $h=0$, the returned signal modes are due solely to the bright noise background (the red cloud).  As such, there is no phase-sensitive cross correlation between the returned signal modes and the stored idler modes, as illustrated by the absence of green dashed lines. At the receiver, the annihilation operators for the signal modes are then $\{\hat{c}_{S_m} = \hat{c}_{N_m} : 1\le m \le M\}$, where the noise modes are now each in thermal states containing $N_B$ photons on average, so that there is no passive signature of target presence~\cite{Tan2008}.  

Conditioned on the true hypothesis $h$, the returned signal and stored idler mode pairs, $\{\hat{c}_{S_m},\hat{c}_{I_m}: 1 \le m \le M\}$, are in iid zero-mean Gaussian states that are completely determined by their Wigner covariance matrices, viz., 
\begin{eqnarray}
& 
{\mathbf{\Lambda}}_h =
\frac{1}{4}
\left(
\begin{array}{cccc}
(2N_B+1) {\mathbf I}&2\sqrt{\kappa N_S(N_S+1)}\,{\mathbf Z}\delta_{1h}\\
2\sqrt{\kappa N_S(N_S+1)}\, {\mathbf Z}\delta_{1h}&(2N_S+1){\mathbf I}
\end{array} 
\right),
&
\label{CovReturnIdler_main}
\end{eqnarray}
for $h=0,1$.  In this expression: ${\mathbf I} = {\rm diag}(1,1)$; ${\mathbf Z}={\rm diag}(1,-1)$;  $\delta_{ih}$ is the Kronecker delta function; we have used $(2N_B + 1)$ in lieu of $(2\kappa N_S + 2N_B + 1)$ because $\kappa \ll 1$, $N_S \ll 1$, and $N_B \gg 1$; and $\sqrt{\kappa N_S(N_S+1)}$ is the residual phase-sensitive cross correlation between the returned signal and stored idler modes that heralds  target presence. It follows that the task of QI target detection is identifying the presence of that phase-sensitive cross correlation.  

At this juncture it is useful to present two generic approaches to sensing whether the returned-signal, stored-idler mode pairs possess a phase-sensitive cross correlation:  local operations plus classical communication (LOCC), and collective operations.  These approaches---shown schematically in Fig.~\ref{FIG_LOCC}---will appear later in the guises of the OPA receiver and the FF-SFG receiver.  For now we merely note the following points.  The LOCC scheme performs unitary transformations followed by positive operator-valued measurements (POVMs) on each mode pair that may be connected by classical (feedforward or feedback) communication and whose outcomes are pooled to determine its decision, $\tilde{h} = 0$ or 1, as to whether the target is absent ($\tilde{h} = 0$) or present ($\tilde{h} = 1$).  The collective approach, in contrast, applies a unitary transformation to \emph{all} of the mode pairs and then performs a single POVM  to generate its $\tilde{h}$.  

\section{The OPA Receiver}
\label{OPArcvr}
\begin{figure}
\centering
\includegraphics[width=0.4\textwidth]{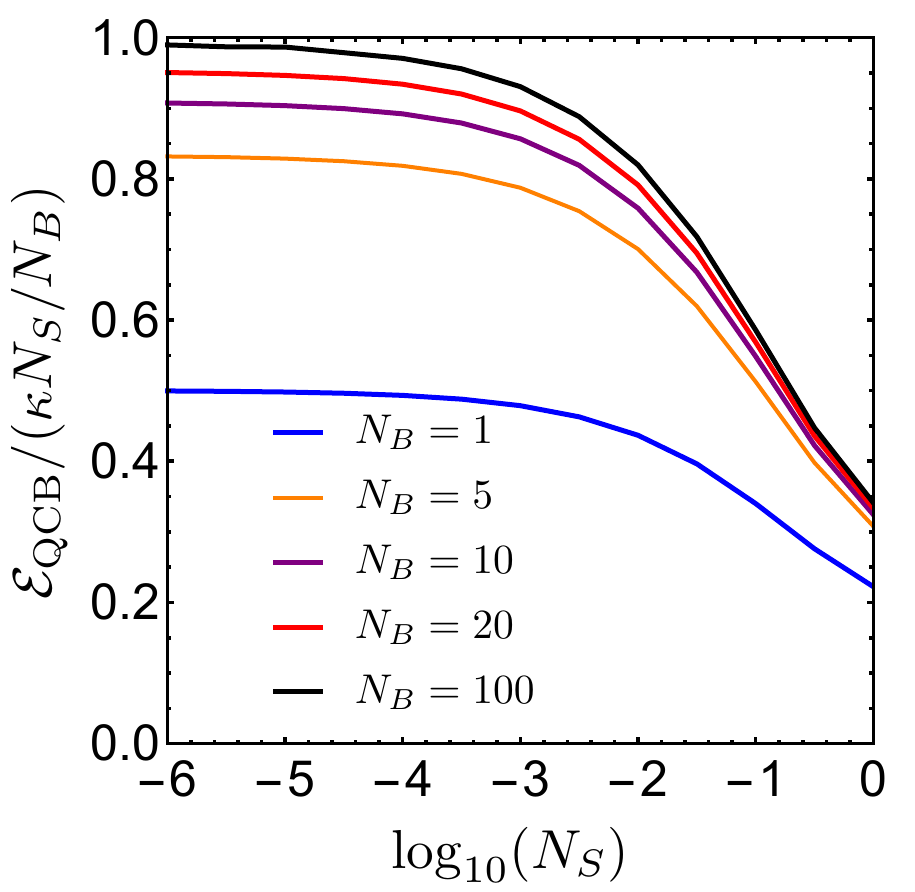}
\caption{
Plots of QI target detection's QCB error-probability exponent normalized by its $N_S \ll 1$, $N_B \gg 1$ asymptote versus $N_S$ for various $N_B$ values. \label{FIG_qcb}
} 
\end{figure}

Helstrom~\cite{Helstrom1969} showed that Neyman-Pearson optimum hypothesis testing---for discriminating between the density operators $\hat{\rho}_h^{\otimes M}$ for $h=0,1$---is realized by taking $\tilde{h}$ to be the outcome of the POVM $u(\hat{\rho}_1^{\otimes M} -\zeta\hat{\rho}_0^{\otimes M})$, where $u(x) = 1$ for $x \ge 0$ and 0 otherwise.  Here, because $P_F = {\rm Tr}[\hat{\rho}_0^{\otimes M}u(\hat{\rho}_1^{\otimes M} -\zeta\hat{\rho}_0^{\otimes M})]$, the constant $\zeta$ is chosen to saturate the Neyman-Pearson criterion's constraint on that quantity.  Unfortunately, analytical expressions for $P_F$ and $P_D= {\rm Tr}[\hat{\rho}_1^{\otimes M}u(\hat{\rho}_1^{\otimes M} -\zeta\hat{\rho}_0^{\otimes M})]$ are unavailable for QI target detection's density operators.  It is worth noting, in this regard, that Helstrom's optimum POVM for minimum error-probability QI target detection takes the form given above with $\zeta = \pi_0/\pi_1$, where $\pi_h$ is the prior probability of hypothesis $h$. 
Helstrom's minimum error-probability POVM leads to an error-probability exponent---$\mathcal{E} \equiv -\lim_{M\to \infty}[\ln(\Pr(e))/M]$, where $\Pr(e) = \pi_0P_F + \pi_1P_M$---that is given by the quantum Chernoff bound (QCB)~\cite{Audenaert2007},
\begin{equation}
\mathcal{E}_{\rm QCB} =  -\ln\!\left[\min_{0\le s\le 1} {\rm Tr}\!\left(\hat{\rho}_0^s \hat{\rho}_1^{1-s}\right)\right],
\end{equation}
for all nondegenerate priors ($\pi_0\pi_1\neq 0$).  Because $\hat{\rho}_0$ and $\hat{\rho}_1$ for QI target detection are both Gaussian states, $\mathcal{E}_{\rm QCB}$ can be obtained analytically~\cite{Pirandola2008,Spedalieri2014}.  It then turns out that $\mathcal{E}_{\rm QCB} \rightarrow \kappa N_S/N_B$ in the $N_S\ll1, N_B\gg1$ limit~\cite{Tan2008}. Figure~\ref{FIG_qcb} plots $\mathcal{E}_{\rm QCB}/(\kappa N_S/N_B)$ versus $N_S$ for a variety of $N_B$ values.  We see that the asymptotic formula works well when $N_B>20$ and $N_S\le 10^{-3}$. 
 
When Ref.~\cite{Tan2008} was published, it was known that none of the three conventional optical receivers---heterodyne, homodyne, or direct detection---yielded any advantage in error-probability exponent in QI target detection.  It was not until the work of Guha and Erkmen~\cite{Guha2009}, and the subsequent experiment by Zhang \emph{et al}~\cite{Zheshen_15}, that an architecture---the OPA receiver---which afforded QI target detection a performance advantage over CI target detection was proposed, analyzed, and experimentally demonstrated.  

\begin{figure}
\centering
\includegraphics[width=0.4\textwidth]{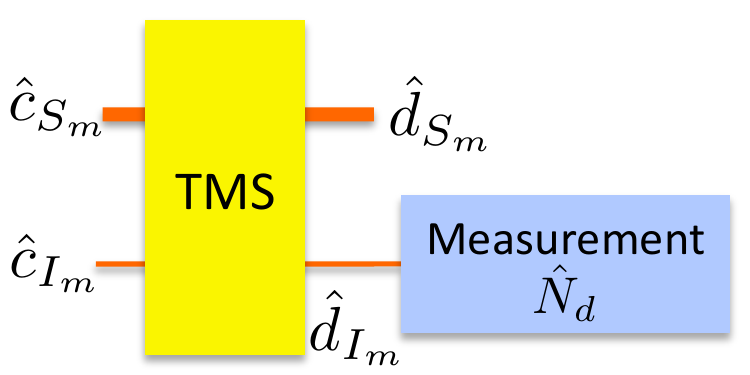}
\caption{Schematic of the OPA receiver~\cite{Guha2009}.
\label{OPA_schematic}
}
\end{figure}

Figure~\ref{OPA_schematic} shows a schematic of the OPA receiver.  Each returned-signal and stored-idler mode pair undergoes a two-mode squeezing (TMS) operation governed by the gain-$G$ Bogoliubov transformation
\ba
\hat{d}_{S_m}&=&\sqrt{G}\hat{c}_{S_m}+\sqrt{G-1} \hat{c}_{I_m}^\dagger
\\
\hat{d}_{I_m}&=&\sqrt{G}\hat{c}_{I_m}+\sqrt{G-1} \hat{c}_{S_m}^\dagger,
\ea
with $0 < G-1 \ll 1$.  This TMS operation converts the absence or presence of a phase-sensitive cross correlation between the $\hat{c}_{S_m}$ and $\hat{c}_{I_m}$ modes into a difference in the average photon number in the $\hat{d}_{I_m}$ mode.  The OPA receiver's $\tilde{h} = 0$ or 1 decision is made by measuring the total photon number in the $\{\hat{d}_{I_m}\}$ modes---i.e., measuring $\hat{N}_d \equiv \sum_{m=1}^M\hat{d}_{I_m}^\dagger\hat{d}_{I_m}$---and comparing its outcome $n_d$ with a threshold that maximizes the posterior probability, viz.,
\begin{equation}
\tilde{h}_{\rm OPA} = \arg\max_j\!\left[\pi_jP_{\hat{N}_d}^{(j)}(n_d)\right],
\end{equation}
where $P_{\hat{N}_d}^{(j)}(n_d)$ is the conditional probability of getting $n_d$ given that $h=j$.   Note that although we have described the OPA receiver on a mode-pair basis, its $M$ TMS operations can be performed simultaneously using a low-gain optical parametric amplifier, and its total photon-number measurement $\hat{N}_d$ can be accomplished by direct detection, thus enabling a convenient experimental realization~\cite{Zheshen_15}.   

The iid nature of the returned-signal and stored-idler mode pairs, conditioned on target absence or presence, implies that the $\hat{d}_{I_m}$ modes will also be iid given $h$.  Furthermore, $M\gg 1$ then provides a central-limit-theorem justification for a Gaussian approximation to the $P_{\hat{N}_d}^{(j)}(n_d)$ distribution.  Using this approximation we can calculate the error probability for equally-likely target absence or presence, minimize it over the OPA gain, and show that the resulting error-probability exponent is 3\,dB inferior to $\mathcal{E}_{\rm QCB}$ in the asymptotic ($\kappa \ll1, N_S\ll 1, N_B \gg 1$) regime.  We have also used the Gaussian approximation and OPA gain optimization to obtain OPA reception's ROC that we will present and discuss in Sec.~\ref{ROCsec}.

The OPA receiver's suboptimality stems from its being an LOCC system~\cite{Acin_2005}.  The LOCC approach is capable of minimum error-probability quantum reception for multiple-copy, pure-state discrimination, but QI target detection in the $\kappa \ll 1, N_S\ll 1, N_B \gg 1$ operating regime is a multiple-copy, mixed-state discrimination problem for which it is known that a collective measurement is needed to achieve that performance~\cite{Calsamiglia_2010}. Indeed, it has been recently shown that QI target detection using LOCC reception can achieve at most a 3\,dB advantage in error-probability exponent over CI~\cite{Sanz_2017}.

\section{The FF-SFG receiver}
\label{SFGrcvr}
We have just seen that the Helstrom POVM for optimum QI target detection cannot be realized with the LOCC approach.  Instead, a collective measurement is required. In principle, that collective measurement can be implemented by a quantum Schur transform~\cite{Bacon_2006} on a quantum computer.  We have recently introduced the FF-SFG receiver~\cite{Zhuang2017}, and showed it to be the first architecture---short of a quantum computer---whose error-probability exponent for equally-likely target absence or presence achieves QI target detection's 6\,dB advantage over CI in $N_S \ll 1$ low-signal-brightness regime.  The FF-SFG receiver builds on two guiding principles:  (1) that SFG is the inverse of the down-conversion process that generates $M$ modes of two-mode squeezed states from a single-mode coherent-state pump; and (2) the Dolinar receiver~\cite{Dolinar1973} achieves minimum error-probability discrimination between arbitrary coherent-state hypotheses.    As noted in~\cite{Zhuang2017}, the FF-SFG receiver can be adapted to realize Helstrom's Neyman-Pearson POVM $u(\hat{\rho}_1^{\otimes M} -\zeta\hat{\rho}_0^{\otimes M})$ for QI target detection merely by modifying the FF-SFG receiver's Bayesian update rule (see below) to use $\pi_1 =1/(1+\zeta)$ for the prior probability of target presence.    

\begin{figure}
\centering
\includegraphics[width=0.7\textwidth]{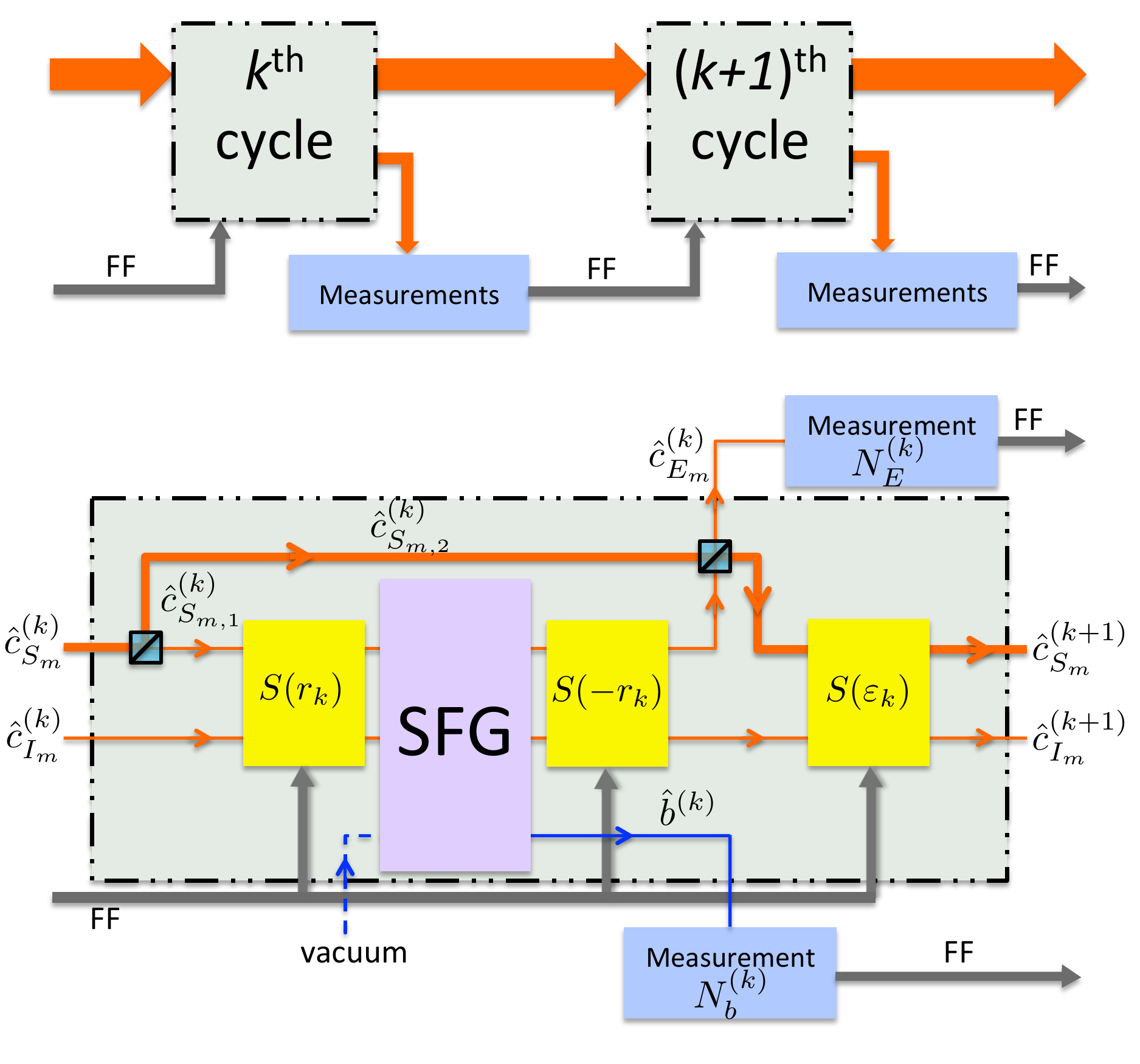}
\caption{Schematic of the FF-SFG receiver~\cite{Zhuang2017} The top panel shows how two of its $K$ cycles are connected.  The bottom panel shows the detail of one of those cycles.  SFG denotes sum-frequency generation.  $S(\cdot)$ denotes a TMS operation. FF denotes feed forward.
\label{SFG_schematic}
}
\end{figure}

The FF-SFG receiver entails $K$ cycles, as shown in Fig.~\ref{SFG_schematic}.  Each cycle employs:  three TMS operations whose squeeze parameters are determined from measurement information fed forward from the preceding cycle; an SFG process that, assuming the previous cycle's tentative decision as to target absence or presence is correct, almost fully converts any phase-sensitive cross correlation in its input modes into auxiliary-mode photons at its output; and photon-number measurements on the auxiliary modes. The photon-number measurement outcomes are fed into a Bayesian-update rule that dictates the next tentative target absence or presence decision based on the information available up to that point in the reception process. The Bayesian-update rule also produces feedforward information that controls the TMS operations in the next cycle.  The total number of cycles is chosen to ensure receiver performance that is close to quantum optimum.   

Now let us explain how the FF-SFG receiver achieves minimum error-probability performance for an arbitrary but given set of priors, $\{\pi_0,\pi_1\}$; for full details see~\cite{Zhuang2017}. (Setting $\zeta = \pi_0/\pi_1$ leads to this receiver's realizing the maximum $P_D$ value consistent with $P_F = {\rm Tr}[\hat{\rho}_0^{\otimes M}u(\hat{\rho}_1^{\otimes M} -\zeta\hat{\rho}_0^{\otimes M})]$.) Akin to the OPA receiver, the FF-SFG receiver converts phase-sensitive cross correlation into photon-number information that can be measured by direct detection. Unlike the OPA receiver, which uses LOCC operations on {\em individual} mode pairs, the FF-SFG receiver applies a joint operation to {\em all} mode pairs. In particular, each of its cycles uses an SFG process that operates on a collection of weak signal-idler mode pairs and a vacuum auxiliary mode~\cite{Zhuang2017}.  If the tentative decision from the previous cycle is correct, this SFG process will convert almost all of any phase-sensitive cross correlation in each signal-idler mode pair into a coherent state of the auxiliary mode embedded in a weak thermal background. Critically, the coherent states that SFG creates from the $M$ mode pairs at its input are in phase.  Thus their coherent-state contributions to the auxiliary-mode output add constructively.  As such, the SFG operation is {\em not} LOCC, opening a path for optimum QI target detection. 

The inputs to the FF-SFG receiver's first cycle ($k=0$) are the returned-signal and the stored-idler mode pairs, represented by annihilation operators $\hat{c}_{S_m}^{(0)} = \hat{c}_{S_m}$ and $\hat{c}_{I_m}^{(0)} = \hat{c}_{I_m}$. A beam splitter with transmissivity $\eta\ll 1$ taps a small portion of each $\hat{c}_{S_m}^{(k)}$ mode, yielding a weak transmitted mode $\hat{c}_{S_m,1}^{(k)}$ to undergo the TMS operation $S(r_k)$ with the $\hat{c}_{I_m}^{(k)}$ mode, and a strong $\hat{c}_{S_m,2}^{(k)}$ mode that is retained. The TMS operation's squeezing parameter, $r_{k}$, is computed from $\tilde{h}_k$, which is the tentative absence or presence decision made prior to the present cycle.  (For the $k=0$ cycle, that tentative decision is derived solely from the prior probabilities.) The $r_k$ value is chosen to almost purge any phase-sensitive cross correlation between the $\{\hat{c}_{S_m,1}^{(k)},\hat{c}_{I_m}^{(k)}\}$ mode pairs from the $S(r_k)$ operation's output mode pairs when $\tilde{h}_k$ is a correct decision~\cite{Zhuang2017}. $S(r_k)$'s output mode pairs undergo an SFG process that converts any residual phase-sensitive cross correlation to photons in the auxiliary sum-frequency $\hat{b}^{(k)}$ mode. Thus, the subsequent detection of photons in the $\hat{b}^{(k)}$ mode is an indication of the tentative decision $\tilde{h}_k$ was incorrect. Following the $k$th cycle's SFG operation, we apply the TMS operation $S(-r_k)$ to each signal-idler mode pair, which ensures that, when its signal-mode outputs are combined with the  retained $\{\hat{c}_{S_m,2}^{(k)}\}$ modes on a second transmissivity-$\eta$ beam splitter, the $\{\hat{c}_{E_m}^{(k)}\}$ output modes contain the same number of photons as the $\hat{b}^{(k)}$ mode.  The photon-number measurements $\hat{b}^{(k)\dagger}\hat{b}^{(k)}$ and $\sum_{m=1}^M\hat{c}_{E_m}^{(k)\dagger}\hat{c}_{E_m}^{(k)}$ then provide outcomes $N_b^{(k)}$ and $N_E^{(k)}$ that are substantial when $\tilde{h}_k$ is incorrect, but negligible when $\tilde{h}_k$ is correct. 
The $k$th cycle is completed by a TMS operation $S(\varepsilon_k)$, with $\varepsilon_k=\sqrt{\eta}\,r_k$,  that makes the phase-sensitive cross correlation of the signal and idler inputs to the $(k+1)$th cycle independent of $r_k$.   

The Bayesian update rule that generates $\{\tilde{h}_k: 0 \le k \le K-1\}$ from the FF-SFG receiver's photon-number measurements works as follows.  For $k=0$, we initialize the process using the given priors in $\tilde{h}_k = \arg\max_j(\pi_j)$.  The prior probabilities for target absence and presence based on all measurement outcomes up to and including those from the $k$th cycle are given by the Bayesian update rule~\cite{Acin_2005,Assalini2011}, 
\be
P_{h=j}^{(k)}=\frac{P_{h=j}^{(k-1)}P_{BE}(N_b^{(k-1)},N_E^{(k-1)};j,r_{\tilde{h}_{k-1}}^{(k-1)})}{\sum_{j=0}^1 P_{h=j}^{(k-1)}P_{BE}(N_b^{(k-1)},N_E^{(k-1)};j,r_{\tilde{h}_{k-1}}^{(k-1)})},
\label{update}
\ee
for $1\le k \le K-1$, where $P_{BE}(N_b^{(k-1)},N_E^{(k-1)};j,r_{\tilde{h}_{k-1}}^{(k-1)})$ is the conditional joint probability of getting counts $N_b^{(k-1)}$ and $N_E^{(k-1)}$ given that the true hypothesis is $j$, $r_{k-1} = r_{\tilde{h}_{k-1}}^{(k-1)}$ is the decision-dependent TMS squeezing parameter for cycle $k-1$, and $P_{h=j}^{(0)} = \pi_j$.  The tentative decision that determines the TMS squeezing parameter for the $k$th cycle is then $\tilde{h}_k = \arg\max_j(P_{h=j}^{(k)})$.   After the last cycle ($k=K-1$), the final decision on target absence or presence is $\tilde{h}_K = \arg\max_j(P_{h=j}^{(K)})$, where the $\{P_{h=j}^{(K)}\}$ are obtained from Eq.~\ref{update} with $k = K$.  This decision accounts for \emph{all} the information obtained from the $K$ measurement cycles.  As we have shown in Ref.~\cite{Zhuang2017}, the total number of cycles needed to approach optimum FF-SFG performance is determined by the beam splitter's transmissivity $\eta$.

\section{Receiver operating characteristic comparison}
\label{ROCsec}
\begin{figure}
\centering
\includegraphics[width=0.35\textwidth]{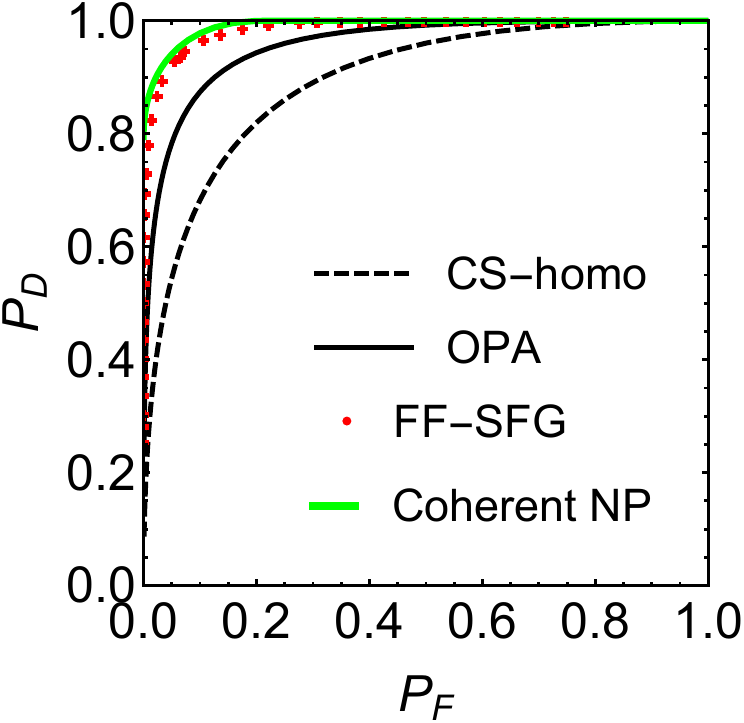}
\caption{
ROCs of QI target detection with FF-SFG reception (red dots), QI target detection with OPA reception (black solid curve), CI target detection with coherent-state (CS) light and homodyne reception (black dashed curve) schemes.  Also included is the ROC of coherent-state Neyman-Pearson (Coherent NP) for discriminating between the coherent state $|\sqrt{M\kappa N_S/N_B}\,\rangle$ and the vacuum state (green solid curve), which is known to be realized by QI target detection with FF-SFG reception when $N_S \ll 1$.  All four ROCs assume that $M = 10^{7.5}$, $N_S=10^{-4}$, $\kappa = 0.01$, and $N_B=20$.
\label{ROC}
}
\end{figure}
The culmination of this paper is the ROC comparison we will present in this section for the $P_D$ versus $P_F$ trade-offs of QI target detection with FF-SFG reception, QI target detection with OPA reception, and CI target detection with coherent-state illumination and homodyne detection.  Also included is the ROC for discriminating between the coherent state $|\sqrt{M\kappa N_S/N_B}\,\rangle$ and the vacuum state, which we have shown in Ref.~\cite{Zhuang2017} to be the FF-SFG's performance when $N_S \ll 1$.  These four ROCs---which are plotted in Fig.~\ref{ROC}---all assumed the same operating parameters:  $M = 10^{7.5}$ transmitted modes, $N_S = 10^{-4}$ average transmitted photon-number per mode; $\kappa = 0.01$ roundtrip channel transmissivity when the target is present; and $N_B = 20$ average received background photon-number per mode.  The FF-SFG receiver's ROC was obtained from Monte Carlo simulations done in the manner described in Ref.~\cite{Zhuang2017}.  In particular, to get a point on the FF-SFG receiver's ROC, we first choose a $\zeta$ value, then initialize the Bayesian update procedure from Eq.~(\ref{update}) using the priors $\pi_0 = \zeta/(1+\zeta)$ and $\pi_1 = 1/(1+\zeta)$, and run the simulation to obtain $P_D$ and $P_F$.  The OPA receiver's ROC was obtained from the Gaussian approximation to its photon-counting statistics conditioned on the true hypothesis, similar to what we have previously done for the use of QI with OPA reception to realize classical communication that is immune to passive eavesdropping~\cite{Zhang2013}.  The coherent-state homodyne setup's ROC was obtained analytically from the conditional statistics of its homodyne receiver's output.  Note that because $N_B \gg 1$, a homodyne receiver is essentially the optimum quantum receiver for CI target detection.  Furthermore, the target-detection problem for the coherent-state homodyne setup reduces to distinguishing between known signals embedded in additive Gaussian noise, whose ROC is well known~\cite{VanTrees1968}.  The ROC for discriminating the coherent state $|\sqrt{M\kappa N_S/N_B}\,\rangle$ from the vacuum state can be obtained analytically, as shown by Helstrom~\cite{Helstrom1976}.  

Figure~\ref{ROC} shows the superiority of FF-SFG reception to OPA reception in QI target detection, and the improvements that both offer over CI target detection. More importantly, Fig.~\ref{ROC} shows that the FF-SFG's ROC in the $N_S\ll 1$ limit matches that of the optimum discrimination between the coherent state $|\sqrt{M\kappa N_S/N_B}\,\rangle$ and the vacuum state, as expected from what was previously found for minimum error-probability QI target detection with equally-likely target absence or presence~\cite{Zhuang2017}.  Thus we conclude that FF-SFG reception provides a structure-receiver alternative to Schur-transform implementation on a quantum computer for achieving the QI's full performance advantage for detecting the presence of a weakly-reflecting target that is embedded in a bright noise background.

\ack
QZ acknowledges support from the Claude E. Shannon Research Assistantship.
ZZ and JHS acknowledge support from AFOSR Grant No.~FA9550-14-1-0052.
QZ thanks M. M. Wilde for helpful discussions.

\section*{References}
\bibliographystyle{iopart-num}

\end{document}